\documentclass{ieeeaccess}
\usepackage{cite}
\usepackage{amsmath,amssymb,amsfonts}
\usepackage{algorithmic}
\usepackage{graphicx}
\usepackage{textcomp}

\usepackage{cleveref} 
\usepackage{subcaption}
\usepackage{url}
\usepackage[binary-units=true]{siunitx}
\usepackage[ruled, lined, commentsnumbered, longend]{algorithm2e}

\def\BibTeX{{\rm B\kern-.05em{\sc i\kern-.025em b}\kern-.08em
    T\kern-.1667em\lower.7ex\hbox{E}\kern-.125emX}}
\begin{document}
\history{}
\doi{}

\crefname{figure}{Fig.}{Figs.}
\crefname{table}{Tab.}{Tabs.}
\crefname{equation}{Eq.}{Eqs.}
\crefname{section}{Sec.}{Sec.}
\Crefname{figure}{Figure}{Figures}
\Crefname{table}{Table}{Tables}
\Crefname{equation}{Equation}{Equations}
\Crefname{section}{Section}{Sections}
\crefname{algorithm}{Algo.}{Algo.}

\title{Simultaneous execution of quantum circuits on current and near-future NISQ systems}
\author{\uppercase{Yasuhiro Ohkura}\authorrefmark{1, 2},
\uppercase{Takahiko Satoh\authorrefmark{1, 3}, and Rodney Van meter}.\authorrefmark{1, 4},
\IEEEmembership{Member, IEEE}}
\address[1]{Keio University Quantum Computing Center, Yokohama, Kanagawa 223-8522 Japan}
\address[2]{Graduate School of Media and Governance, Keio University SFC, Fujisawa, Kanagawa 252-0882 Japan (e-mail: rum@sfc.wide.ad.jp)}
\address[3]{Graduate School of Science and Technology, Yokohama, Kanagawa 223-8522 Japan}
\address[4]{Faculty of Environment and Information Studies, Keio University SFC, Fujisawa, Kanagawa 252-0882 Japan}
\tfootnote{This work was supported by MEXT Quantum Leap Flagship Program Grant Number JPMXS0118067285}

\markboth
{Ohkura \headeretal: Simultaneous quantum circuits execution on current and near-future NISQ systems}
{Ohkura \headeretal: Simultaneous quantum circuits execution on current and near-future NISQ systems}

\corresp{Corresponding author: Yasuhiro Ohkura (email: rum@sfc.wide.ad.jp).}

\begin{abstract}
In the NISQ era, multi-programming of quantum circuits (QC) helps to improve the throughput of quantum computation.
Although the crosstalk, which is a major source of noise on NISQ processors, may cause performance degradation of concurrent execution of multiple QCs, its characterization cost grows quadratically in processor size.
To address these challenges, we introduce palloq (parallel allocation of QCs) for improving the performance of quantum multi-programming on NISQ processors while paying attention to the combination of QCs in parallel execution and their layout on the quantum processor, and reducing unwanted interference between QCs caused by crosstalk.
We also propose a software-based crosstalk detection protocol that efficiently and successfully characterizes the hardware's suitability for multi-programming. 
We found a trade-off between the success rate and execution time of the multi-programming.
This would be attractive not only to quantum computer service but also to users around the world who want to run algorithms of suitable scale on NISQ processors that have recently attracted great attention and are being enthusiastically investigated.
\end{abstract}


\titlepgskip=-15pt

\maketitle

\section{Introduction}
\label{sec:introduction}

\begin{figure*}[ht]
    \centering
    \resizebox{0.95\textwidth}{!}{\includegraphics{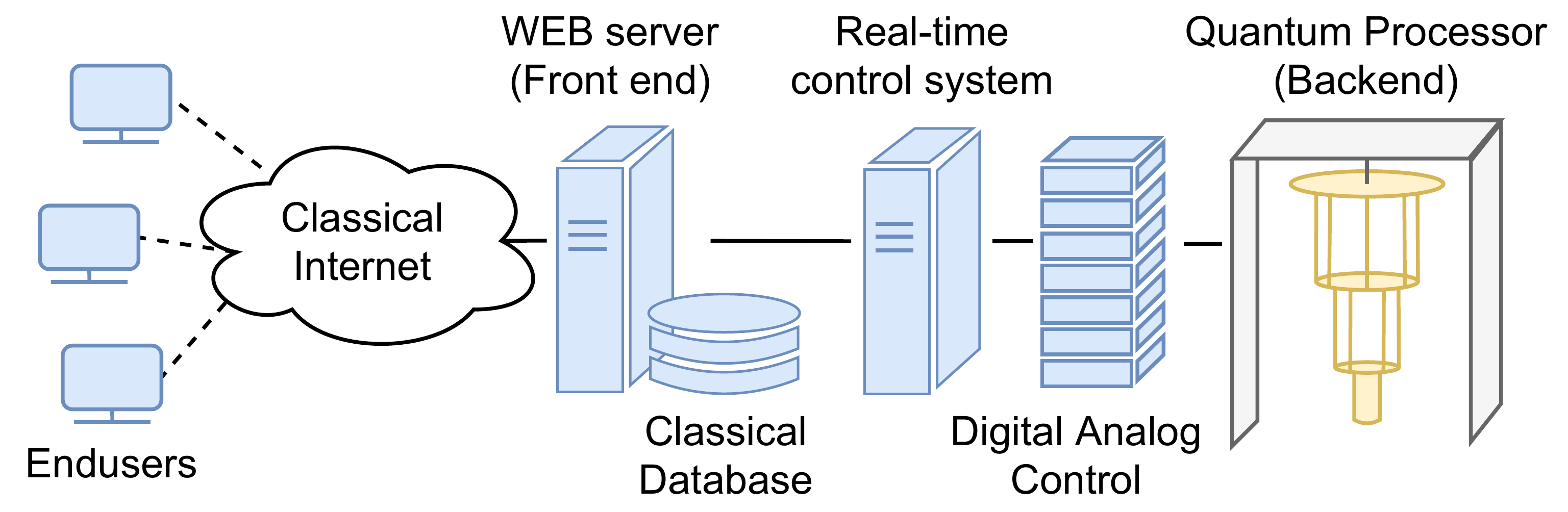}}
    \caption{
        {\bf Cloud Quantum Computing Architecture}
        Various implementations of quantum processors have been proposed \cite{jones2001quantum, clarke2008superconducting, kielpinski2002architecture, dutt2007quantum, o2009photonic}.
        As the core of the service, the quantum processor provides its computing resources by performing quantum operations and measurements.
        Digital Analog control exchanges the quantum and classical information by converting program instruction into analog signals and measurement results into classical data.
        The real-time control system is responsible for classical-quantum interaction, mid-circuit measurements, and feed-forward operations.
        Another end of the cloud computing is the browser-based user interface.
        End-users create the requests (jobs) on a web browser, send them to the system via a web server through the Internet, and receive the results of the computation.
        The browser-based user interface provides AAA (authentication, authorization, accountability), and in some cases, quantum programming tool and its development environment, such as a GUI-based quantum circuit composer.
        }
    \label{fig:cloud_quantum}
\end{figure*}

Current processors, called Noisy Intermediate-Scale Quantum (NISQ) ~\cite{preskill2018quantum}, are not immune to noise, which causes a high error rate and greatly affects the reliability of the computation.

With the advent of cloud quantum computing systems, quantum computing has become familiar to researchers and developers around the world.
The more users and tasks from various demands and backgrounds have increased, the more it is important to maximize the throughput of the NISQ processors.
To operate efficiently, executing multiple quantum tasks concurrently can be one solution.
However this method is not trivial and involves fundamental challenges~\cite{Das2019, liu2020, Ohkura2021, niu2021enabling} .

It is difficult to explain the whole range of errors of computation on NISQ processors by using only information from standard error characterization techniques such as randomized benchmarking~\cite{Knill2008, Chow2009}.

Though quantum tomography estimates statistical reliability of a quantum state, it is requires large amounts of data exponential in the number of qubits for complete state tomography.
To maximize the utilization and performance of a NISQ processor, we should take into account not only the standard model of isolated qubits but also the detection model for context-dependent errors, which requires a small cost ~\cite{Rudinger2019}.

In the case of superconducting qubit systems, effects of crosstalk on the gate errors is a serious problem ~\cite{Mundada2019, McKay2019}.
When multiple quantum circuits (QC) are executed in parallel, as resource usage of the processor increases, unwanted interference may occur due to crosstalk noise between independent QCs, which may affect the calculation results.

In this work, we introduce palloq (parallel allocation of QCs), a system for improving the performance of multi-programming on NISQ processors. 
Our system consists of two parts: first, the multiple circuit composer to maximize searches for an effective combination of the application circuits (a knapsack-like problem); 
second, the crosstalk-aware layout method that allocates hardware qubits to multiple QCs taking into account both local error rate and non-local noise (crosstalk). 
We also take into account the cost of crosstalk characterization and propose a novel detection method. 
We show the performance of our system by executing dozens to low hundreds of queued quantum tasks as a multi-programming workload on the real-world cloud quantum computing platform, IBM Quantum Experience, and measure the success rate of the individual QCs and total execution time.

This paper is organized as follows. 
In \cref{sec:background}, we review the cloud quantum computing environment, quantum multi-programming and crosstalk effect on NISQ processors.
In \cref{sec:crosstalkcharacterization}, first, we describe the crosstalk characterization approach in previous work, then introduce our novel characterization method. 
In \cref{sec:palloq}, we propose compilation methods for efficient multi-programming. 
In \cref{sec:experiments}, we show the experimental results on quantum processors and evaluate our proposed approach.
Finally, we conclude paper in \cref{sec:conclusion}
\section{Background}
\label{sec:background}

\begin{figure*}[ht]
    \begin{subfigure}[c]{0.5\textwidth}
        \centering
        \resizebox{0.9\textwidth}{!}{\includegraphics{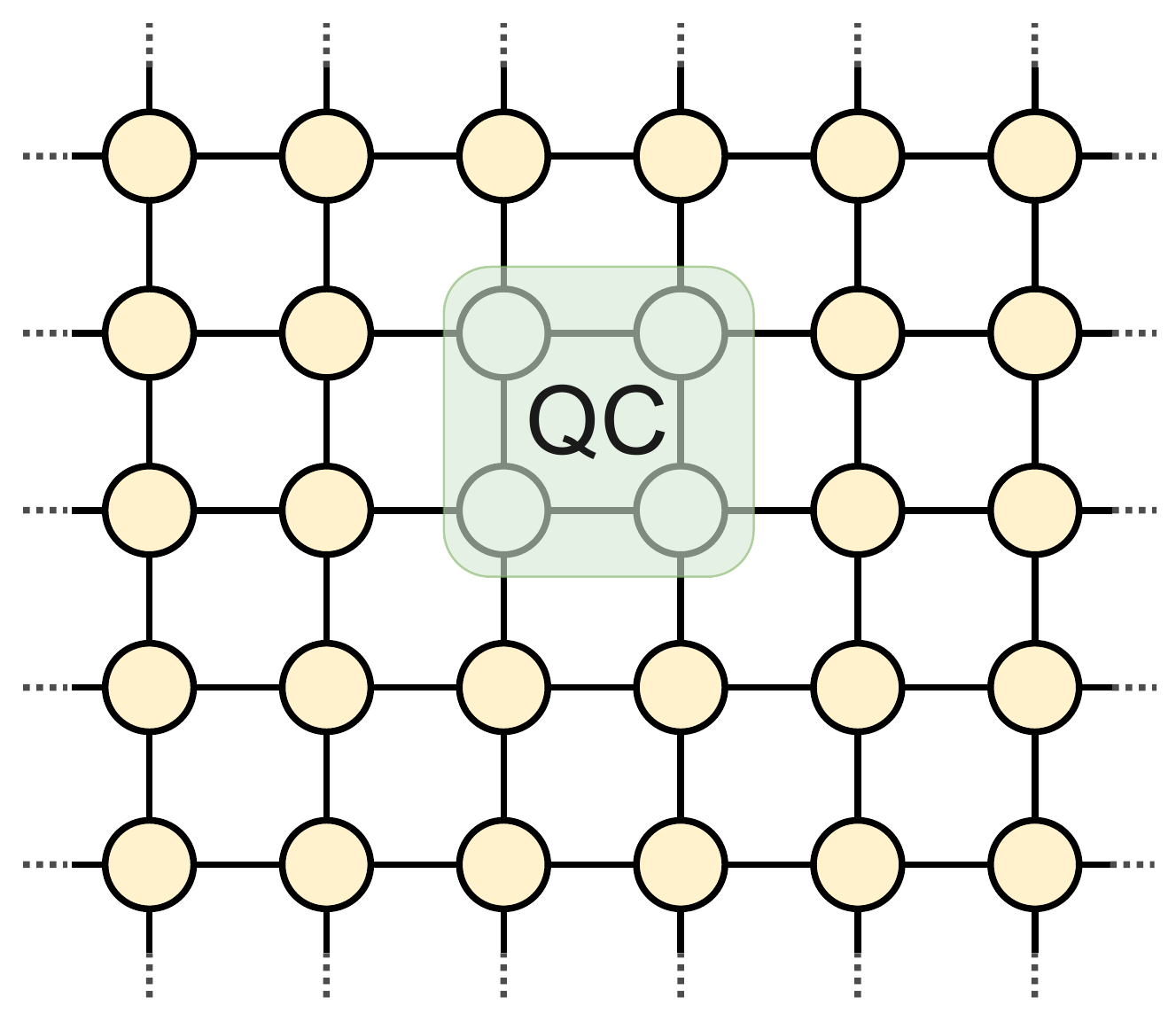}}
        \subcaption{Single execution}
        \label{single_execution}
    \end{subfigure}
    \begin{subfigure}[c]{0.5\textwidth}
        \centering
        \resizebox{0.9\textwidth}{!}{\includegraphics{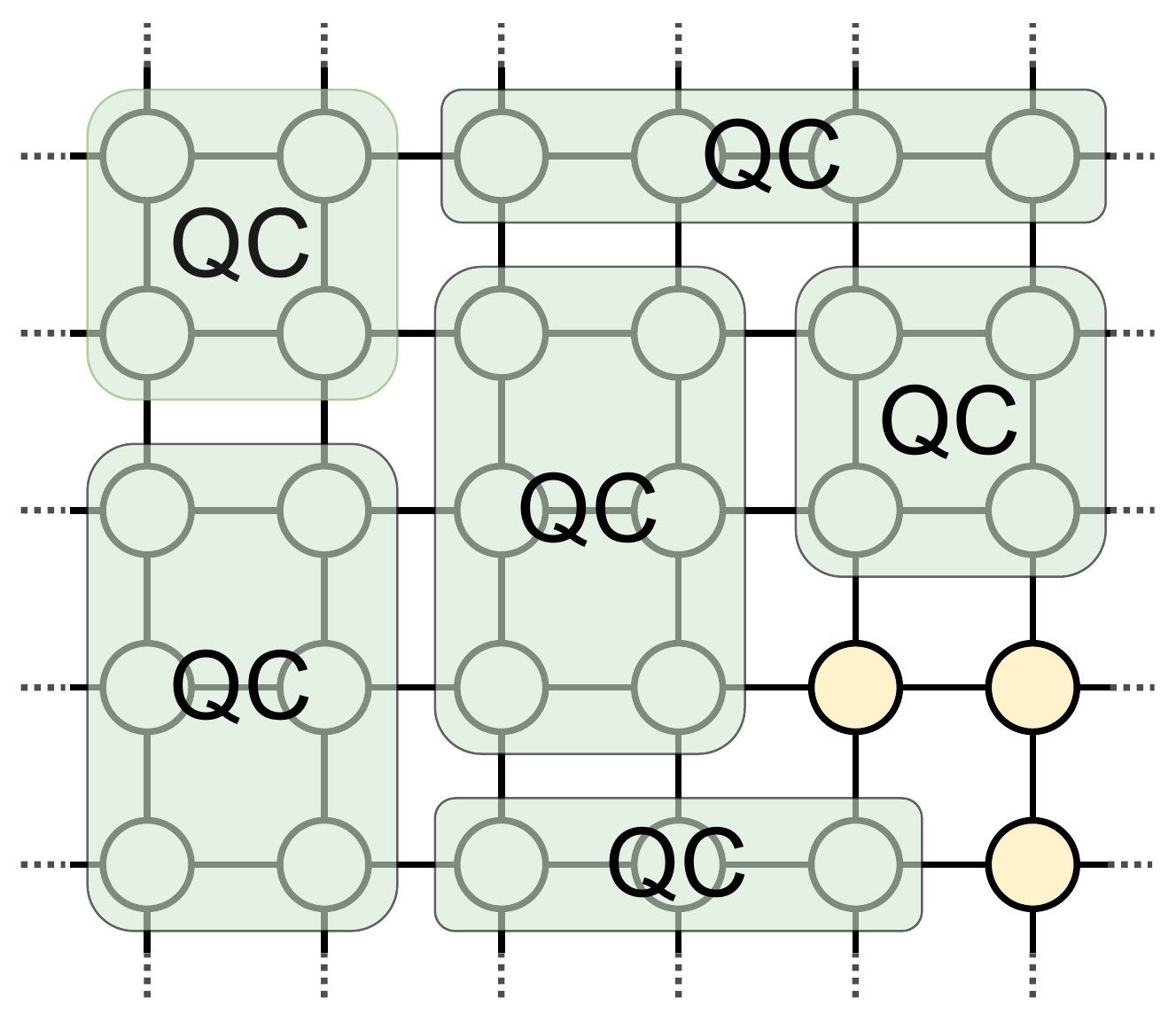}}
        \subcaption{Multiple execution}
        \label{multiple_execution}
    \end{subfigure}
    \caption{
        {\bf The idea of multiple execution. }
        The lattice graph represents the quantum processor, nodes denote qubits and edges are two qubits connection.
        The QC with the green box represents quantum circuits placed on physical qubits of the processor. 
        \cref{single_execution} represents single quantum circuit execution. 
        Because the limited sized circuits are tolerable on NISQ system, the idle qubits (yellow circles) reduce the throughput of the processor. 
        \cref{multiple_execution} describes the idea of multiple circuit execution concurrently.
        This approach reduces idle qubits and is expected to increase the throughput.
    }
    \label{fig:multi_programming}
\end{figure*}

\subsection{Cloud Quantum Computing}

With the advent of cloud quantum systems i.e. quantum computing as a service (QCaaS) \cite{amazonbraket,dwaveleap,googleplayground,ibmqex:2021,microsoftazureq, rigetti,xanadu}, many researchers and developers from a variety of domains are becoming quantum users. 
QCaaS provides the quantum resource that allows opportunities ranging from conducting basic experiments \cite{PhysRevA.94.032329} to developing applications that include quantum simulation, quantum machine learning, and optimization \cite{montanaro2016quantum}.

Cloud quantum computing architectures consist of components that include Quantum Processors, Analog Digital control, Real-time control systems, WEB servers and classical databases, the Internet, and End users, as in  \cref{fig:cloud_quantum}.

Why use a cloud quantum system instead of a local server or a quantum laptop?
In general, there are several reasons for the migration of the service from desktop and corporate server rooms to a cloud platform \cite{acm_cloud_computing, acm_view_of_cloud}.
For the individual users, the total control of the software, OS and low-level utility, and subsequent revisions to other programs comes with a price.
For the service providers, the internet-based service can be developed, tested, and operated on the platform provider's choice instead of coping with various user's environments.
In the case of quantum computation, the development and operation, which includes daily calibration, of a quantum computer are very expensive and  specialized tasks \cite{tannu2018case, mckay2018qiskit}.
Languages, tools, and environments for the development of the quantum program are still not sufficient.
By providing them comprehensively as cloud services, users can utilize the quantum resources without being bothered by maintenance.

The rapid increase in users, urgent access for limited quantum resources, and the number of queued jobs are becoming serious issues. 

\subsection{Performance of Quantum Computer}
Several metrics for performance analysis of quantum computers have been proposed \cite{PhysRevLett.127.100501, lubinski2021application, wack2021quality}. 
Quantum Computing performance is governed by three factors: 1). the size of the problem that be encoded which is determined by the number of physical qubits on a processor, 2). the size of a quantum circuit that can be faithfully executed, which is mainly determined by error rates of each operation and lifetime of qubits, and 3). the number of circuits that can be executed per unit time, which is related to the quantum and classical processing speed.
In this paper, we focus on the improvements of 2: the output fidelity of QCs, and 3: the number of QCs executed at a time by a quantum multi-programming.

\subsection{Quantum Multi-programming}
Quantum multi-programming is a method for improving the throughput and utilization of the NISQ processor by executing multiple QCs simultaneously, instead of keeping the unemployed qubits idle, as shown in  \cref{fig:multi_programming}. 
In previous work, several challenges were discussed as follows \cite{Das2019,liu2020}: 
1). Fair hardware resource allocation for every individual task. 
The difficulty of this issue comes from the variations of characteristics of each physical qubits in the processor including operational error rates and qubit lifetimes \cite{10.1145/3297858.3304007}. 
To solve this, the compiler needs to take this error information into account to optimize the circuit. 
2). Avoidance of unwanted interference between the individual QC.
3). Optimization of the operational timing of each circuit to minimize the unnecessary decoherence effect. 
In the case of multi-programming several QCs with different depth (duration of execution), the shorter circuits suffer wait duration until the longer circuit's operation ends, which may cause the decaying of a quantum state prepared by shorter circuits and reduce the output reliability. 

Improving the utilization of the processor by executing multiple programs concurrently can increase the unwanted interference between independent QCs. 
To reduce serious destructive interference, one option is to monitor and compare the performance of parallel execution and to feed the result into the next execution phase, either single or in parallel \cite{Das2019}.
Rather than that, Ohkura\cite{Ohkura2021}, and Dou and Liu \cite{niu2021enabling} discussed directly focusing on the crosstalk noise on the device which causes non-local errors on QCs of multi-programming. 
They tried to characterize the crosstalk in the processor and optimize qubit allocation along with it.
The problem is the crosstalk characterization grows quadratically in the number of hardware qubits, as we discuss in \cref{sec:crosstalkcharacterization}.

\subsection{Crosstalk in NISQ processor}
\label{sec:crosstalkinnisq}
Crosstalk is known to be a significant source of noise in the quantum processor. 
This type of error can be explained from several aspects, but it is simply the unwanted interaction between coupled qubits in the processor.
It is known that there is a trade-off between the strength of qubit interaction and the magnitude of unwanted crosstalk noise \cite{Gambetta2012,Sheldon2016}.
One type of crosstalk is caused by simultaneous operations between specific pairs of qubits. 
In this paper, we focused on the unwanted interaction due to the two qubit (CX gate) operations.
This types of crosstalk is known to occur in the current quantum architectures including superconducting systems and trapped ions \cite{Krantz_2019,Ospelkaus2008}.

The tuning and mitigation of crosstalk directly become big challenges when developing larger processors \cite{Sheldon2016,rudinger2019probing,Harper2020}. 
There are several software approaches to reducing crosstalk error introduced in previous work. 
In the case of tunable quantum processors including Google's architecture\cite{arute2019quantum}, we can tune qubit frequencies or control specific couplers to disable and shut down the leakage errors \cite{Mundada2019,ding2020systematic}.
In contrast, in fixed frequency qubit systems including IBM Q System, we can optimize the circuit scheduler to avoid concurrent execution of correlated qubits in the processor \cite{Murali2020}.
In this paper, we focused on this fixed qubit system, and provide the solution by a novel layout method in the circuit compilation process.
\section{Crosstalk Characterization}
\label{sec:crosstalkcharacterization}

To understand the performance of quantum processors affected by local and non-local errors and its ability to  concurrently execute of multiple QCs, we need to characterize crosstalk in the processor. 
To simplify the problem, we only take into account how big the average crosstalk on the processor is rather than where these occur, i.e. the location of qubits.

\subsection{The cost of characterization}

\begin{figure}[t]
    \centering
    \resizebox{0.5\textwidth}{!}{\includegraphics[]{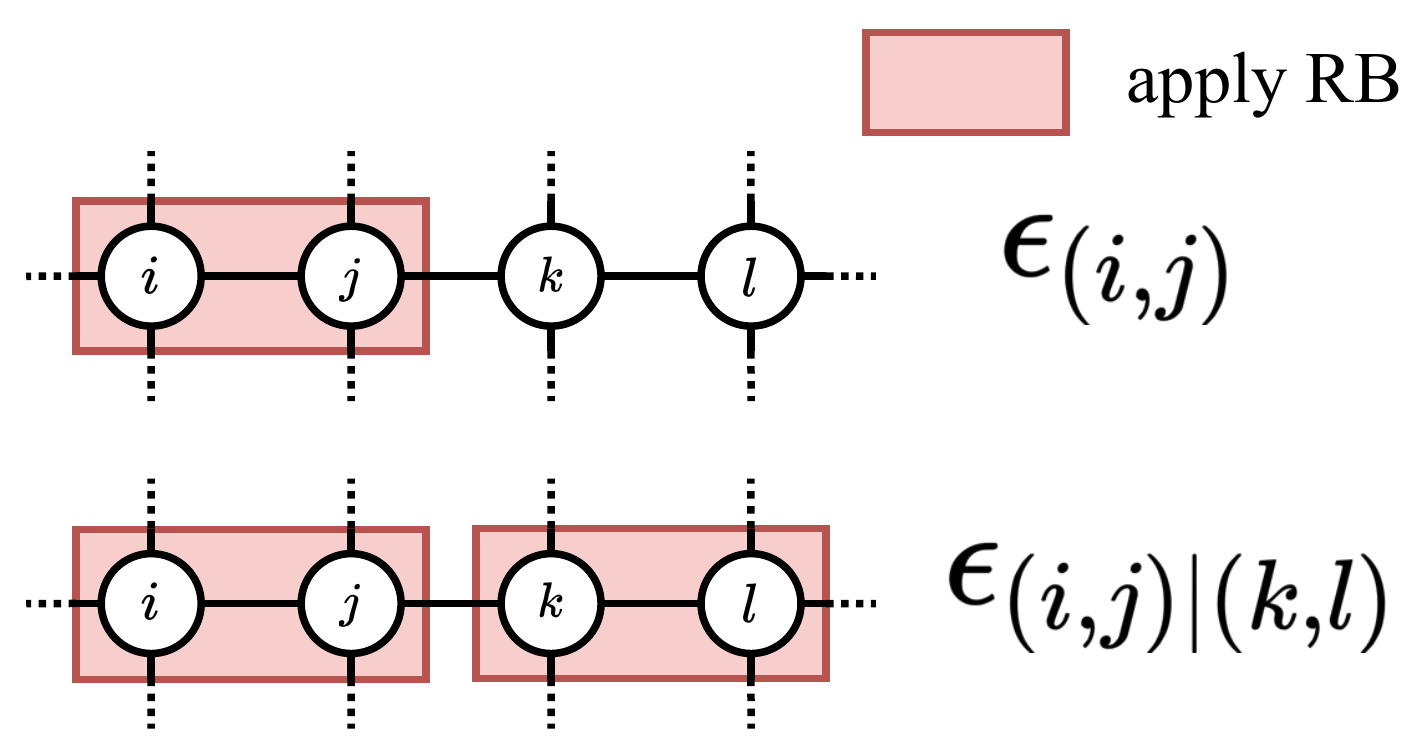}}
    \caption{
        {\bf Simultaneous Randomized Benchmarking (SimRB)}
        Upper diagram shows ordinary RB on two-qubits. 
        RB applies random clifford operations with varying the length of gates and estimates its error rates. 
        In the case of Simultaneous RB, applying RB on more than two hardware areas and comparing the error rates to single RB case, measure the conditional error rates of hardware qubits.
        In this research, we only conducted two-qubits RB and SimRB. 
    }
    \label{fig:simrb_ex}
\end{figure}

The complexity of crosstalk characterization often scales exponentially with the system size.
Recently, several works showed ways to suppress the cost of crosstalk characterization.
One practical protocol introduced in \cite{Murali2020, niu2021enabling} is the comparison of the error rate in the case of individual and parallel execution by using Simultaneous Randomized Benchmarking (SimRB). 
For example, in \cref{fig:simrb_ex}, some pairs of two-qubit errors can be detected in parallel, e.g. ($q_i$, $q_j$) and ($q_k$, $q_l$). 
If the error rate from SimRB $\epsilon ( q_i, q_j ) \vert (q_k, q_l )$ is significantly different from the individual RB $\epsilon ( q_i, q_j )$, there is an unwanted correlation between them.

The combination of these two-qubit pairs grows quadratically in the size of the processor. 
To avoid this situation, in a previous study \cite{Murali2020}, Murali \emph{et~al.} provide some rules to reduce this overhead.
1). Characterize one hop pairs.  
Through the experiments, they found the tendency of occurrence of crosstalk is limited only at one hop on the IBM Q system they used.
This rule is suppressing the detection cost by ignoring the pairs more than one hop apart. 
They also pointed out some older devices have long-range crosstalk strong enough to be a concern.  
2). Characterize high crosstalk pairs only.
They also found the existence of crosstalk tends to be stable in time and space. 

Although this method can detect crosstalk distribution on the processors, it still takes several hours, and it may be impractical in the current situation because the size of processors is continuously getting bigger, and these experiments are queued and run on the cloud system in the presence of other participants.

\subsection{Physical buffer and success rate}
\label{sec:qcdistance}
\begin{figure}[ht]
    \centering
    \resizebox{0.47\textwidth}{!}{\includegraphics{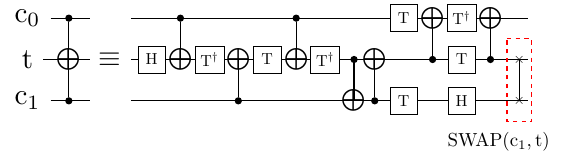}}
    \caption{
        {\bf Toffoli gate on chain topology}
    }
    \label{fig:toffoli_transpiled}
\end{figure}

In this section, we introduce a physical buffer, which is the number of idle qubits between QCs, to mitigate crosstalk effect of the concurrent execution of multiple QCs.
We conducted a preliminary experiment to quantify how the physical buffer affects the output reliability of individual tasks in the concurrent execution.
We used the Toffoli gate as benchmark.
Due to the compilation, that includes the SWAP gate that consists of three CX gates and the qubit routing along with the topology of the processor, shown in \cref{fig:toffoli_transpiled}, for a total of 10 CX gates. 

\begin{figure}[ht]
    \begin{subfigure}[c]{0.5\textwidth}
        \centering
        \resizebox{\textwidth}{!}{\includegraphics{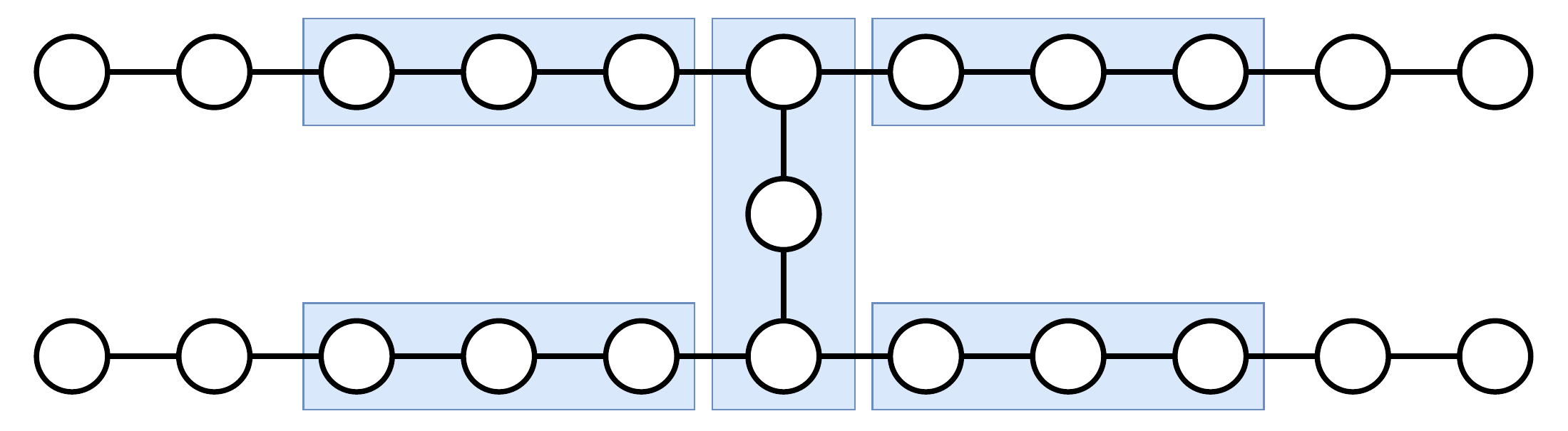}}
        \caption{0 physical buffer}
        \label{0_hop}
    \end{subfigure}
    
    \begin{subfigure}[c]{0.5\textwidth}
        \centering
        \resizebox{\textwidth}{!}{\includegraphics{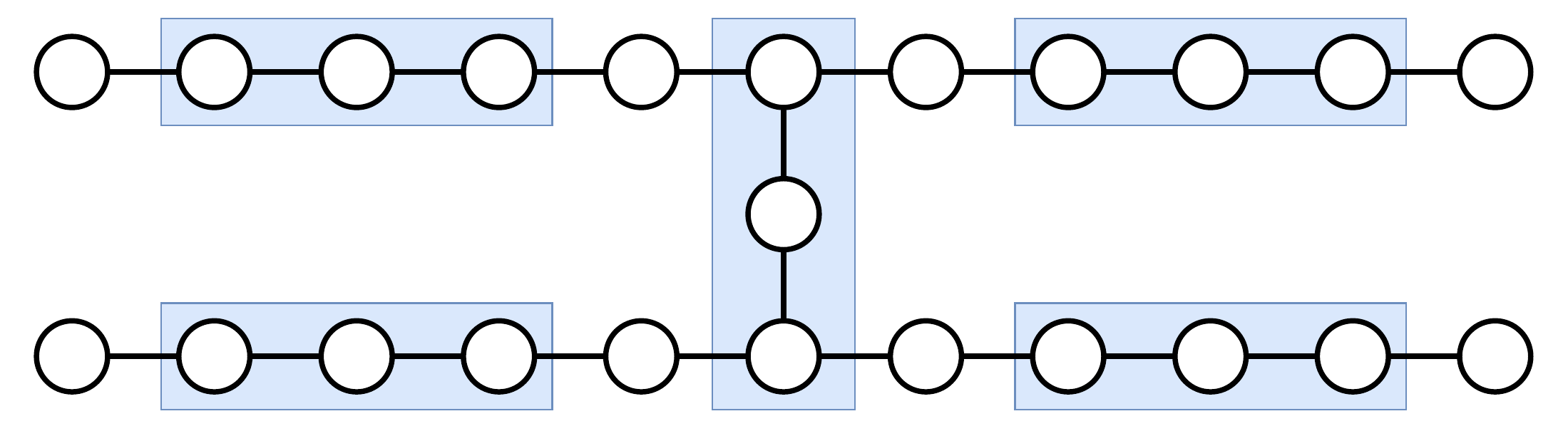}}
        \caption{1 physical buffer}
        \label{1_hop}
    \end{subfigure}
    
    \begin{subfigure}[c]{0.5\textwidth}
        \centering
        \resizebox{\textwidth}{!}{\includegraphics{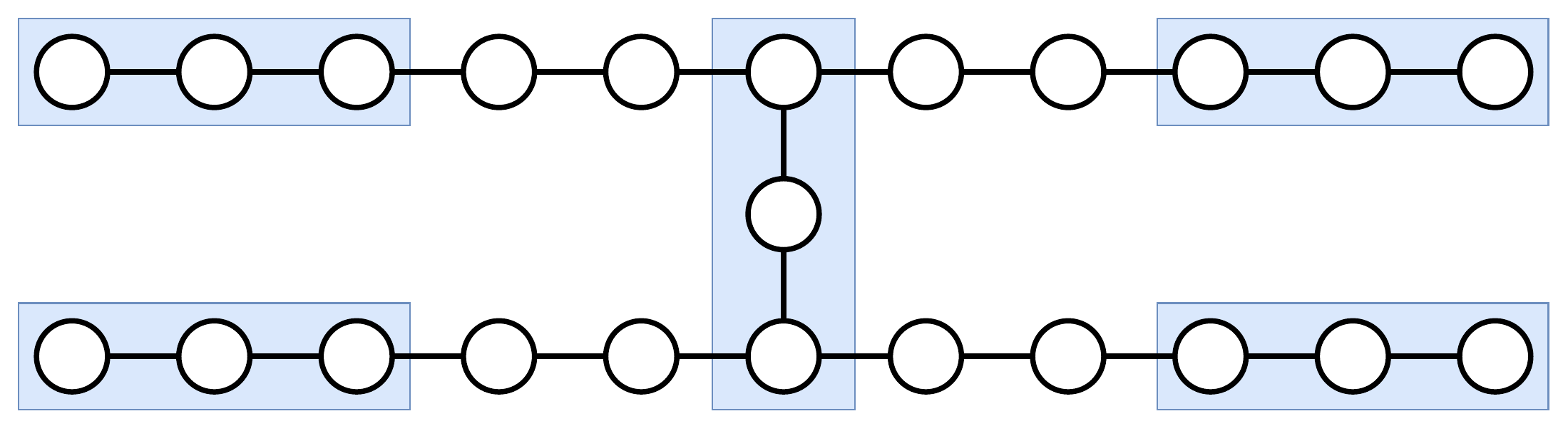}}
        \caption{2 physical buffer}
        \label{2_hop}
    \end{subfigure}
    
    \caption{
        {\bf Multiple Toffoli placement varying physical buffer. }
        The graph denotes the quantum processor, nodes are qubits and edges are two qubit connection of the superconducting qubit system.
        The blue boxes represent the quantum circuits of Toffoli operation placed on physical qubits. 
        We placed 5 circuits and vary the physical buffer, then measure the success rate of the Toffoli placed in the center.
    }
    \label{fig:toffoli_buffered}
\end{figure}

\begin{figure}[ht]
    \centering
    \resizebox{0.5\textwidth}{!}{\includegraphics{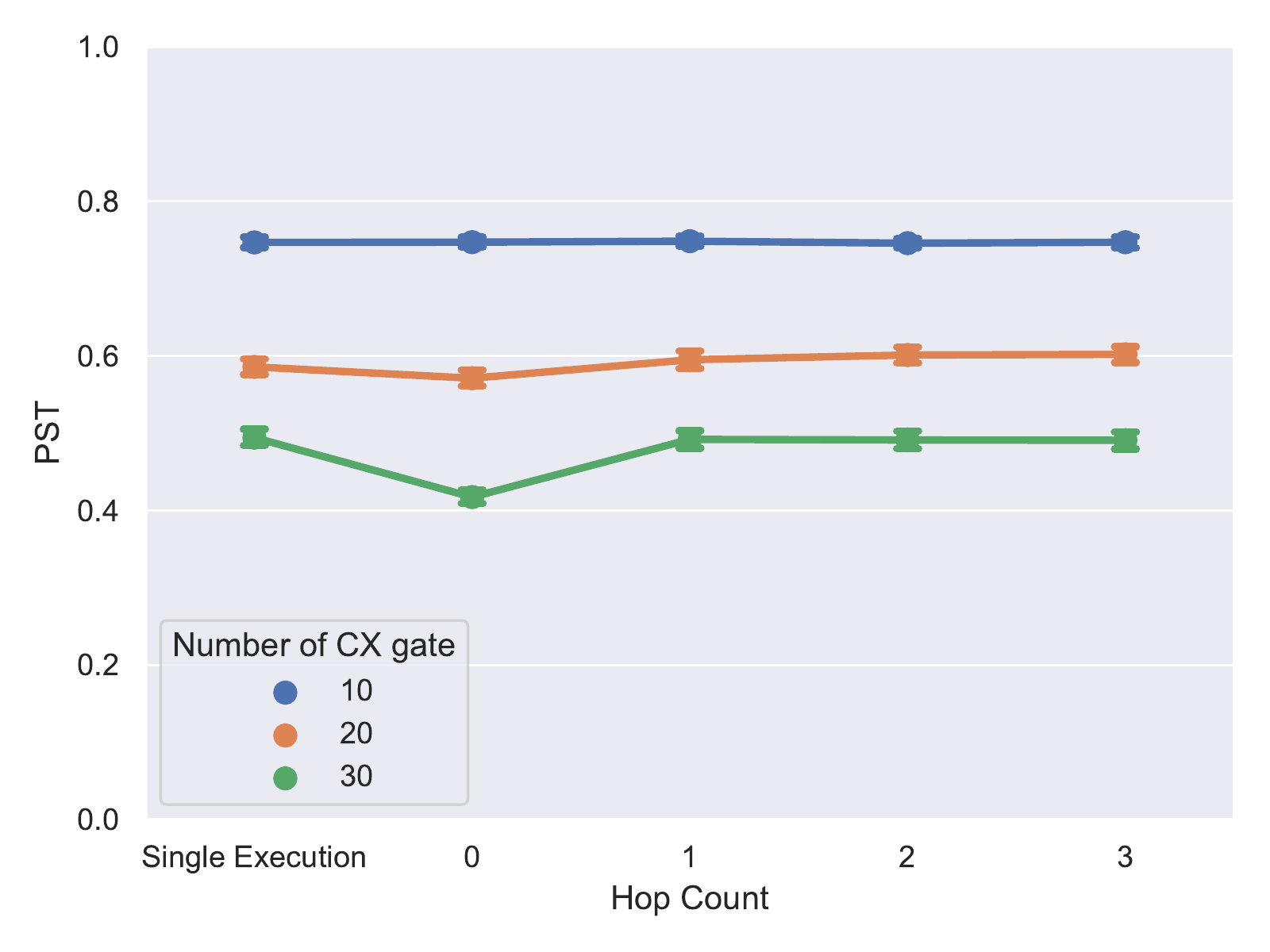}}
    \caption{
        {\bf Physical buffer of multiple circuits and Success rate}
        Each line denotes Probability of Successful Trial (PST) \cref{pst} of Toffoli gates in the case of concurrent execution.
        The horizontal axis is the physical qubit buffer introduced in \cref{fig:toffoli_buffered}. 
        We compared the PST to the left most data point, which is the single execution case. 
        Each color describes the number of CX gates contained in the operation, 10, 20 and 30. 
        Since we utilize the Toffoli gates with SWAP \cref{fig:toffoli_transpiled}, one toffoli gate contains 10 CX gates. 
    }
    \label{fig:qc_buffered_manhattan}
\end{figure}

We varied the number of physical buffer between circuits and the number of Toffoli gates as parameters to see how the success rates change, shown in \cref{fig:toffoli_buffered}. 
\cref{fig:qc_buffered_manhattan} shows that in only the case of adjacent circuits, i.e. no physical buffer, the success rate significantly drops.
With the increase of number of buffer, the success rate is recovered.
And this change appears only for circuits of more than $20$ CX gates.
This leads us to the following insights.
For the concurrent execution in practice, 1). we don't care about crosstalk for the shorter circuits, and 2). there is a threshold number of hops of physical buffer that can improve the output fidelity. 
And in this case, 1 physical buffer is enough until the CX circuit depth reaches $30$.


\subsection{Benchmark of crosstalk immunity}

Although crosstalk is the major source of error and may decrease the reliability of the concurrent execution, with the increasing number of qubits in the processors, the cost to characterize the noise increase quadratically.
We show a novel detection protocol combining RB methods with relatively low detection costs. 
To analyze the crosstalk tolerance of processors, we utilized the coefficient of variation of gate errors as the metric and compared several processors.

We focus on how the crosstalk impacts the average and variance of error rates.
First, we apply RB for every qubit or two-qubit pair in the processor and calculate the average and variance of error rates in the single execution case. 
Then, we run SimRB for all the qubits at the same time and also calculate the average and variance. 
Comparison of average and variance of error rates between those two cases leads to the quantitative analysis of crosstalk effects on the whole performance and immunity of the processor. 

\begin{figure}[ht]
    \centering
    \resizebox{0.5\textwidth}{!}{\includegraphics{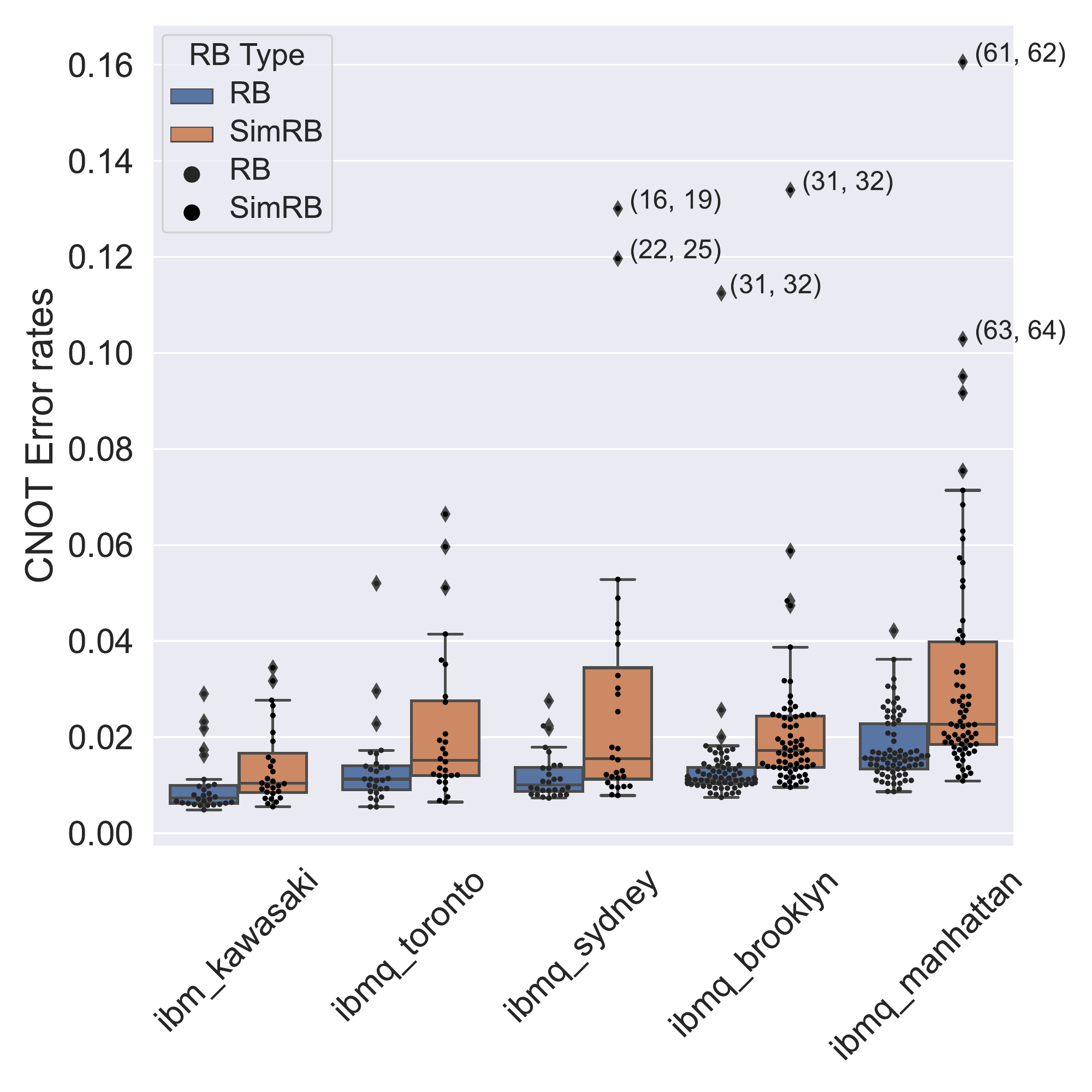}}
    \caption{
        {\bf Crosstalk immunity}
        Each black dot represents CX gate error rate of two qubit pairs.
        $(i, j)$ denotes the label of two qubit pair which is remarkably high error rate.
    }
    \label{fig:crosstalk_immunity_IBMQ}
\end{figure}

We conducted this benchmark on several current IBM quantum processors and showed the comparison of those performances in \cref{fig:crosstalk_immunity_IBMQ}.
We measured the CX gate error rate of each processor by utilizing RB. 
Blue box plots represent the distribution of CX gate error rate and black dots are the error value of each physical two-qubit connection.
The orange box plots also represent error rates but in the case of concurrent execution (SimRB).
We ran several patterns of CX gates combination that can be executed concurrently and took the average value of error rates of each case.

For all processors, the variance and average error rates increase in the case of concurrent execution, indicating the presence of the crosstalk noise.
In particular, IBMQ Toronto, IBMQ Sydney, and IBMQ Manhattan show significant interference by other operations performed on the other regions.
The performance of multi-programming directly depends on this crosstalk interference as we discuss in \cref{sec:experiments}. 

\section{The palloq System}
\label{sec:palloq}

We proposes \emph{palloq}, a system including layout synthesis for multiple QCs and a job scheduler to manage efficient and high fidelity quantum multi-programming. 
The detail of procedure are explained in \cref{sec:pseudocode} Pseudocode.
We published the source code at \url{https://github.com/rum-yasuhiro/palloq}. 

The palloq responsible for the compilation phase and the user's job management which is provided by WEB server in the cloud computing architecture we showed in \cref{fig:cloud_quantum}.

This compiler pass takes several QCs written in OpenQASM \cite{cross2017open}   and the local gate error information of device as input.

Our layout synthesis consists of a heuristic based on Noise-Adaptive layout which analyzes the device's calibration data and searches for better allocation using a greedy approach \cite{murali2019noise}. 
First, it parses the calibration data and hardware qubit connection to create a weighted graph $G_{HW}(V, E)$.
$V$, $E$, and weight represents the physical qubits as the vertices, two-qubit connections of the hardware as the edges, and the reliability $r = 1 - \epsilon$, where $\epsilon$ is error rate of two-qubit operation between physical qubits.
In the same way, each input QC is treated as a weighted directed graph $G_{QC}(V, E)$, where vertices $V$ are the qubits in the QC, edges $E$ are two-qubit gates, and weight is the number of two-qubit gates performed on the same two-qubit pair. 
The compiler searches for the best reliable physical qubits candidates heuristically and allocates them to the highest weighted edges in the circuit's graph. 
Repeat this procedure until all the circuit components are placed. 

As we discussed in \cref{sec:qcdistance}, for the shorter circuits, we don't care about the crosstalk, and for the relatively larger circuits, we only care about the physical distance between circuits and optimize them locally. 
Our software takes physical qubit distance as input. 
Every time each circuit is allocated to hardware qubits, then disable the qubits around them to create a distance to others.
\section{Experiments and Evaluation}
\label{sec:experiments}

To evaluate our proposal, we conducted an experiment varying the physical buffer among the multiple circuits.
In the entire experiment, we focus on the output reliability and total execution time and hardware usage. 

We use small benchmark circuits from previous work \cite{li2020qasmbench} as shown in Table \ref{tab:benchmark}.
The details of the processors and the software we used are shown in Table \cref{tab:software}.

\begin{figure*}[ht]
    \centering
    \resizebox{\textwidth}{!}{\includegraphics{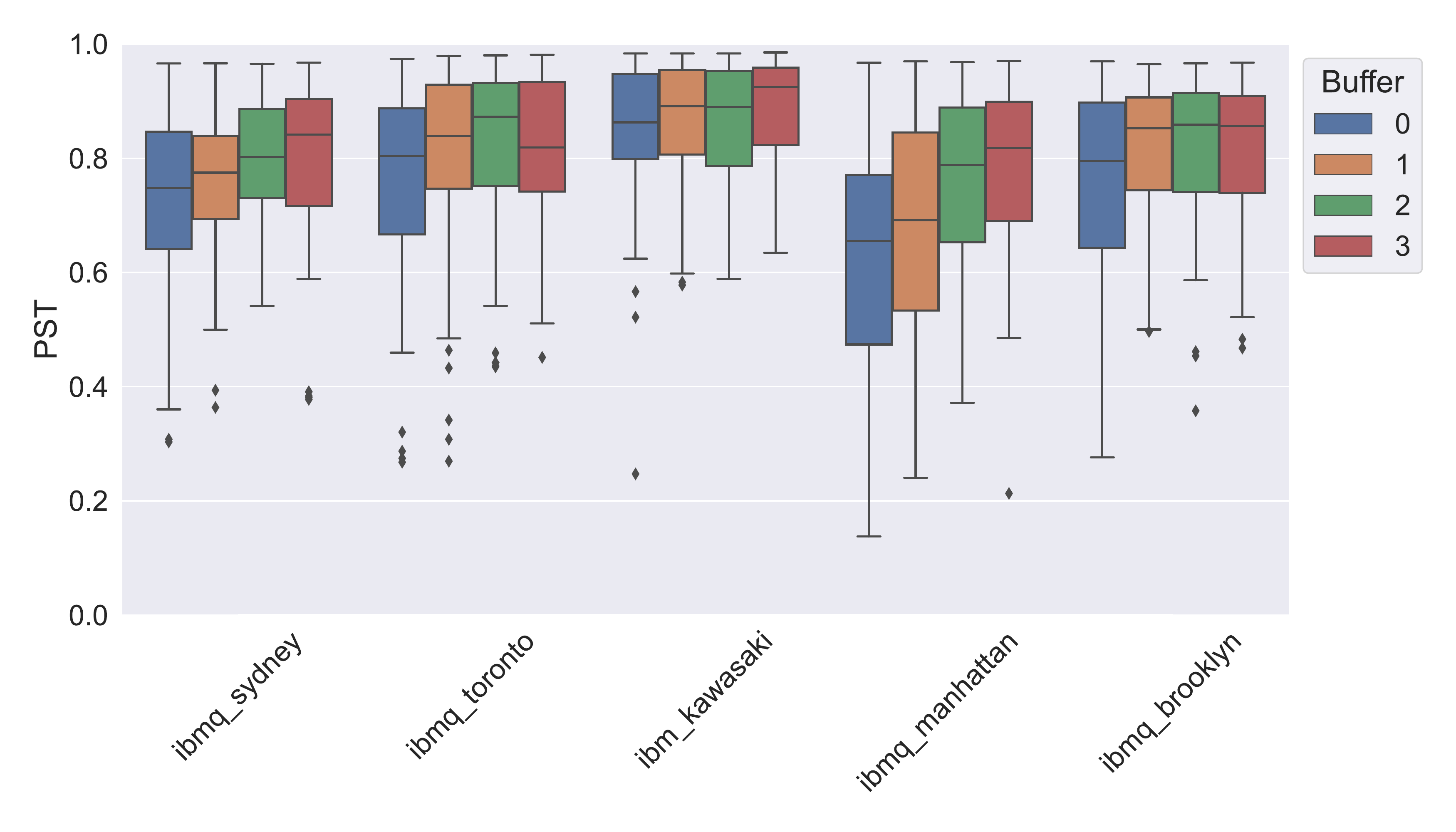}}
    \caption{
        {\bf Success rate and the physical buffer}
        These box plots represent the distribution of PST of benchmark circuits executed on the real quantum devices.
    }
    \label{fig:buffer_and_pst}
\end{figure*}
\begin{figure}
    \centering
    \resizebox{0.48\textwidth}{!}{\includegraphics{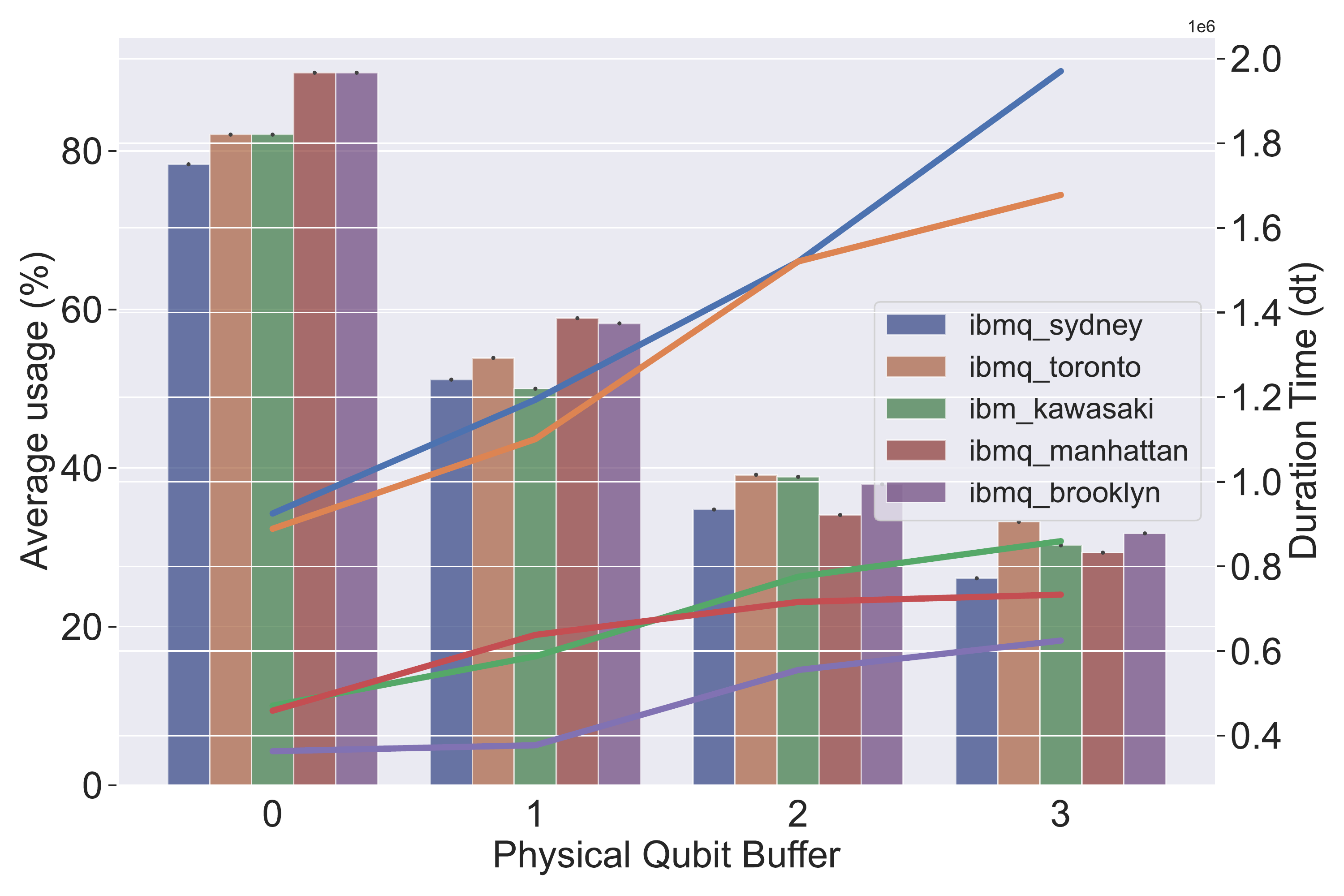}}
    \caption{
        {\bf Hardware usage and total execution time}
    }
    \label{fig:usage_and_time}
\end{figure}

\subsection{Metrics}

To quantify the performance of our proposed method, we utilized Probability of Successful Trial (PST) \cite{10.1145/3297858.3304007}. 
For NISQ system that run the given programs a specified number of times, known as "shots", the quantum computation outputs the answer as a probability distribution of candidate bit strings.
The number of successful trials is defined as how many times the each shots hit the correct bit strings.
We chose the benchmark circuits whose computational result includes only one or a small number of answers for the solution space for ease of handling with PST.
When the QC finishes successfully without any error, PST is 1.

As a baseline we compare the results from a noisy quantum device to a classical simulator as a noiseless case. 
PST is defined as follows:
\begin{equation}
    \rm PST = \frac{\rm Number\ of\ Successful\ Trials}{Number\ of\ Total\ Trials}
    \label{pst}
\end{equation}

\subsection{Execution duration and output reliability}

First, we prepared 100 QCs from the benchmark set and queued them as an input to our compiler. 
Varying the physical qubit buffer 0 to 3, we count the output PST and total execution time of concurrent execution. 

\cref{fig:buffer_and_pst} shows that PST of 100 circuits varies with physical buffer.
For all quantum devices, in the case of buffer more than one, the average PST is better than the case of buffer zero, which is the densest layout and highest through put case. 
Which means, in these experiments, we obtain higher reliability of computation at the expense of throughput of concurrent execution.

\cref{fig:usage_and_time} shows total execution time of 100 circuits as we vary physical buffer. 
For the execution time, we count circuit duration time ($dt$) of all concurrent execution round and $1\ dt = \frac{2}{9}\ ns$. 
It shows a larger buffer reduces hardware usage and total execution time linearly. 
For all devices, around 80 \% of physical qubits are used in the case of buffer 0. 
In the case of buffer 2 and 3, the hardware usage is only half of the densest case. 
On the other hand, for all devices, the total circuit duration time increased by a factor of two from the densest to sparsest.

\subsection{Analysis}

Here we describe the trade-off between crosstalk and throughput of concurrent execution.
In general we desire the success of computation with higher reliability and throughput.
The success rate of concurrent computation, which depends on the gate fidelity (local errors) and crosstalk.
The throughput, which is defined as the number of executed QC per $dt$, where $dt$ is duration time of quantum computation, relies on many factors like gate duration time, measurement duration time, and the size of processors.
Based on experimental results shown in \cref{fig:buffer_and_pst},  we defined the improvement (gain) of output reliability $g$ as:

\begin{equation}
    g =  max(\overline{\rm P S T}_i - \overline{\rm P S T}_j)
    \label{gain_pst}
\end{equation}

where the $\overline{\rm PST_{i}}$ is average value of PSTs of $i$ physical qubit buffered layout of concurrent execution.
In this experiment, for all cases, $i = 0$ and $j$ is $2$ or $3$.
Based on the result shown in \cref{fig:crosstalk_immunity_IBMQ}, we defined the crosstalk presence in the processor $ct$ as:
\begin{equation}
    ct = CV_{\rm SimRB} - CV_{\rm RB}
    \label{ct}
\end{equation}
utilizing the coefficient of variation (CV): 
\begin{equation}
    CV = \frac{\sigma}{\mu}
    \label{cv}
\end{equation}
where $\sigma$ is the standard deviation and $\mu$ is the average of the given distribution.
In this case, crosstalk presence $ct$ represents the degree of change of variation of CX error rate from RB to SimRB.

\begin{figure}[ht]
    \centering
    \resizebox{0.5\textwidth}{!}{\includegraphics{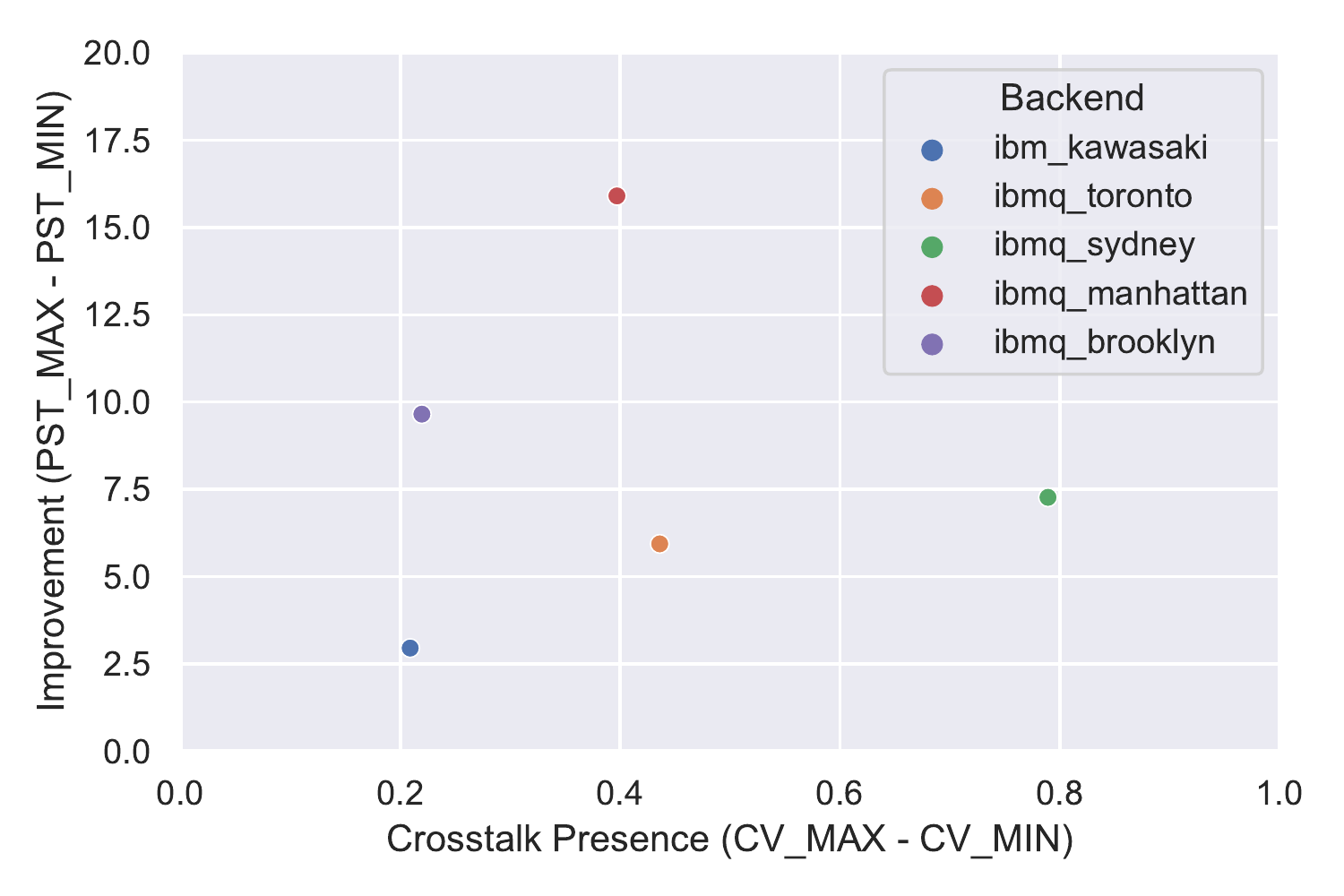}}
    \caption{
        {\bf Crosstalk presence and improvement of concurrent execution}
    }
    \label{fig:ct_imp}
\end{figure}

\cref{fig:ct_imp} shows the relationship between the presence of crosstalk on the device and gain of improvement by physical buffer. 

We used three 27-qubit devices and two 65-qubit devices, and for both cases we can see the positive correlation.
Referring to \cref{fig:usage_and_time}, as we increase the number of physical qubit buffer, the throughput of computation drops.
For processors with relatively strong crosstalk, it is worth reducing the throughput to gain the improved output fidelity of the computation.

\section{Conclusion}
\label{sec:conclusion}

This paper proposed and evaluated a compiler method for concurrent execution of multiple quantum circuits which includes layout and schedule.
First, we showed a practical crosstalk characterization method to reduce the detection cost that is critical for the near-term scale quantum processors. 
For the evaluation, we show our compiler efficiently processes the multiple quantum circuits avoiding the crosstalk and trade-off between the success rate of quantum circuits and the throughput of the processors.

Also taking into account the crosstalk noise is meaningful not only for improving the performance of multi-programming but also when considering the security of future cloud quantum computation.
\appendix
\section{Setup for experiments}

The quantum circuit for benchmark our proposed software, listed in \cref{tab:benchmark}.
The processors we use are listed in \cref{tab:processors}
And the version of software packages we use are listed in \cref{tab:software}

\begin{table*}
    \centering
    \begin{tabular}{l l c c c}
    \hline
        Benchmark & Description & Qubits & Gates & CX \\ \hline
        deutsch & Deutsch algorithm with 2 qubits for $f(x) = x$ & 2 & 5 & 1  \\
        grover & Grover’s algorithm & 2 & 16 & 2  \\ 
        linearsolver & Solver for a linear equation of one qubit & 3 & 19 & 4  \\ 
        toffoli & Toffoli gate & 3 & 18 & 6 \\ 
        fredkin & Fredkin gate & 3 & 19 & 8 \\ 
        adder & Quantum ripple-carry adder & 4 & 23 & 10 \\
        error\_correctiond3 & Error correction with distance 3 and 5 qubits & 5 & 114 & 49 \\ \hline
    \end{tabular}
    \caption{
        {\bf Small Benchmark Circuits. }
        We picked several small size quantum circuit to benchmark our proposal from the benchmark circuit set called QASMBench \cite{li2020qasmbench}.
    }
    \label{tab:benchmark}
\end{table*}
\begin{table*}[ht]
    \begin{tabular}{c c}
        \begin{minipage}[t]{0.59\textwidth}
            \begin{center}
                
\begin{tabular}{l|c}
    \hline
    Name & Spec \\ \hline
    Intel Core i$9$ processor & $\SI{2.3}{\giga\hertz}$,  
    $\SI{32}{\giga\byte}$ RAM \\
    IBMQ Toronro & Falcon r4 $27$-qubit QV32\\
    IBMQ Sydney & Falcon r4 $27$-qubit QV32\\
    IBM Kawasaki & Falcon r5.11 $27$-qubit QV32\\
    IBMQ Manhattan & Hummingbird r2 $65$-qubit QV32\\
    IBMQ Brooklyn & Hummingbird r2 $65$-qubit QV32\\
    \hline
\end{tabular}
\caption{{\bf Processors}}
\label{tab:processors}
            \end{center}
        \end{minipage}
        \begin{minipage}[t]{0.4\textwidth}
            \begin{center}
                \begin{tabular}{l|c}
    \hline
    Name & Version \\ \hline
    Python & $3.8.5$ \\
    qiskit & $0.29.0$ \\
    qiskit-terra & $0.17.0$ \\
    qiskit-aer & $0.8.2$ \\
    qiskit-ignis & $0.5.1$ \\
    qiskit-ibmq-provider & $0.16.0$ \\ \hline
\end{tabular}
\caption{{\bf Software version}}
\label{tab:software}
            \end{center}
        \end{minipage} \\
    \end{tabular}
\end{table*}

Each experiment was conducted on the dates listed in \cref{tab:experiments}.
\begin{table*}[ht]
    \centering
    \begin{tabular}{l|c}
        \hline
        Experiment & Date-time \\ \hline
        Physical buffer of multiple circuits in \cref{fig:qc_buffered_manhattan} on IBMQ Manhattan & 2021/05/10 ~ 2021/05/14 \\ \hline
        
        Crosstalk detection in \cref{fig:crosstalk_immunity_IBMQ} on IBMQ Toronto & 2021/07/09 \\ \hline
        
        Crosstalk detection in \cref{fig:crosstalk_immunity_IBMQ} on IBMQ Sydney & 2021/07/09 \\ \hline
        
        Crosstalk detection in \cref{fig:crosstalk_immunity_IBMQ} on IBM Kawasaki & 2021/07/09 \\ \hline
        
        Crosstalk detection in \cref{fig:crosstalk_immunity_IBMQ} on IBMQ Manhattan & 2021/07/10 \\ \hline
        
        Crosstalk detection in \cref{fig:crosstalk_immunity_IBMQ} on IBMQ Brooklyn & 2021/07/10 \\ \hline
        
        Performance analysis of proposal compiler in \cref{fig:buffer_and_pst} on IBMQ Toronto & 2021/08/25 \\ \hline
        
        Performance analysis of proposal compiler in \cref{fig:buffer_and_pst} on IBMQ Sydney & 2021/08/25 \\ \hline
        
        Performance analysis of proposal compiler in \cref{fig:buffer_and_pst} on IBM Kawasaki & 2021/08/25 \\ \hline
        
        Performance analysis of proposal compiler in \cref{fig:buffer_and_pst} on IBMQ Manhattan & 2021/08/25 \\ \hline
        
        Performance analysis of proposal compiler in \cref{fig:buffer_and_pst} on IBMQ Brooklyn & 2021/08/25 \\ \hline
        
    \end{tabular}
    \caption{{\bf Date and time when experimental data have been taken}}
    \label{tab:experiments}
\end{table*}


\section{Pseudocode}
\label{sec:pseudocode}
Here we show the pseudocode of the algorithm we introduced in \cref{sec:palloq}.

\SetKwComment{Comment}{/* }{ */}
\SetKwInput{KwIn}{Input}
\SetKwInOut{KwOut}{Output}

\begin{algorithm*}
\label{code:distance_layout}
\caption{Physical distance layout}
\KwIn{Queued DAG circuits $\bf queue$,  Processor's topology graph weighted by CX gate reliability $\bf G_{HW}$, Buffer to adjacent circuit $\bf d$}
\KwOut{List of Multi-programming circuits with layout as \bf MP}

\

Set {\bf MP} to empty list\;

${\bf queue} = sort\{\bf queue\  \vert\  \bf CX\  depth\}$\;

$S := \{{\bf G_{HW_{S}}} \in {\bf G_{HW}}\ \vert\ {\bf connected\ subgraph\ of}\ {\bf G_{HW}}\}$\;

\While{$ length({\bf queue}) > 0$}{
    
    ${\bf DAG} = {\bf queue}.pop(0)$\;    
    
    \eIf{$size(max(S)) > size({\bf DAG})$}{
        
        Set {\bf layout}, {\bf pending\_qubits}, {\bf prog}, {\bf hw} to empty list\;
        Set {\bf multi\_QC} to empty DAG\;
        
        ${\bf G_{QC}}\ \underleftarrow{\bf convert}\ {\bf DAG}$\Comment*[r]{Convert {\bf DAG} to graph ${\bf G_{QC}}$ weighted by number of CX gates}

        ${\bf pending\_qubits} \leftarrow {\bf G_{QC}}.{\rm nodes}$\; 
        
        ${\bf q_{prog1}}, {\bf q_{prog2}} \leftarrow max\{{\bf G_{QC}}.{\rm edges}\}$\Comment*[r]{Heaviest program edge}
        ${\bf pending\_qubits}.remove({\bf q_{prog1}}, {\bf q_{prog2}})$
        
        ${\bf prog}.append({\bf q_{prog1}}, {\bf q_{prog2}})$\;
        
        ${\bf q_{hw1}}, {\bf q_{hw2}} \leftarrow max\{{\bf G_{HW_{S}}}.{\rm edges}\ \vert \ size({\bf G_{HW_{S}}})>size({\bf DAG})\}$\Comment*[r]{Most reliable HW qubits}
        
        \While{${\bf pending\_qubits}\ not\ null\ set$}{
            
            ${\bf q_{prog}} \leftarrow {\bf q_{adj}} \in max\{{\rm edge}({\bf q_{adj}}, {\bf q_{in}}) \in {\bf G_{QC}}\ \vert \ {\bf q_{in}} \in {\bf prog}\}$\;
            
            ${\bf q_{hw}} \leftarrow {\bf q_{adj}} \in max\{{\rm edge}({\bf q_{adj}}, {\bf q_{in}}) \in {\bf G_{HW_{S}}}\ \vert \ {\bf q_{in}} \in {\bf hw}\}$\;
            
            \
            
            ${\bf pending\_qubits}.remove({\bf q_{prog}})$
            
            \
            
            ${\bf layout}.append(\{{\bf q_{prog}}, {\bf q_{hw}}\})$\;
            ${\bf prog}.append({\bf q_{prog}})$\;
            ${\bf hw}.append({\bf q_{hw}})$\;
        }
        
        Delete used qubits\;
        ${\bf G_{HW}}.remove({\bf hw})$\;
        ${\bf G_{HW}}.remove({\bf {q} \in {\bf G_{HW}}}\ \vert\ {\rm {\bf d}-neighbored\ to\ } {\bf hw})$\;
        
        $S := \{{\bf G_{HW_{S}}} \in {\bf G_{HW}}\ \vert\ {\bf connected\ subgraph\ of}\ {\bf G_{HW}}\}$\;
        
        ${\bf multi\_QC}.append({\bf DAG})$
    }{
        ${\bf MP}.append(\{ {\bf multi\_QC}, {\bf layout} \})$\;
        Reinitialize hardware graph ${\bf G_{HW}}$\;
        $S := \{{\bf G_{HW_{S}}} \in {\bf G_{HW}}\ \vert\ {\bf connected\ subgraph\ of}\ {\bf G_{HW}}\}$\;
    }
}
\Return{\bf MP}
\end{algorithm*}

\section*{Acknowledgment}

\bibliographystyle{IEEEtran}
\bibliography{references}

\EOD

\end{document}